\documentstyle[12pt]{article} 
\input psfig.sty

\def\Title#1{\begin{center} {\Large #1 } \end{center}}
\def\Author#1{\begin{center}{ \sc #1} \end{center}}
\def\Address#1{\begin{center}{ \it #1} \end{center}}

\def\doeack{\footnote{Work supported by the Department of Energy,
                     contract DE--AC03--76SF00515.}}
\def\SLAC{Stanford Linear Accelerator Center\\
    Stanford University, Stanford, California 94309 USA}

\newenvironment{Abstract}{\begin{quotation} \begin{center}
                       ABSTRACT
     \end{center}\bigskip  }{\end{quotation}}
\def\beq{\begin{equation}}
\def\eeq#1{\label{#1}\end{equation}}
\def\eeqn{\end{equation}}
\def\beqa{\begin{eqnarray}}
\def\eeqa#1{\label{#1}\end{eqnarray}}
\def\eeqan{\end{eqnarray}}

\def\Re{{\cal R \mskip-4mu \lower.1ex \hbox{\it e}\,}}
\def\Im{{\cal I \mskip-5mu \lower.1ex \hbox{\it m}\,}}
\def\nn{\noindent}
\def\ie{{\it i.e.}}
\def\eg{{\it e.g.}}
\def\etc{{\it etc}}
\def\etal{{\it et al.}}

\def\sub#1{_{\lower.25ex\hbox{$\scriptstyle#1$}}}
\def\sul#1{_{\kern-.1em#1}}
\def\sll#1{_{\kern-.2em#1}}  
\def\sbl#1{_{\kern-.1em\lower.25ex\hbox{$\scriptstyle#1$}}}
\def\ssb#1{_{\lower.25ex\hbox{$\scriptscriptstyle#1$}}}
\def\sbb#1{_{\lower.4ex\hbox{$\scriptstyle#1$}}}

\def\mh{\ifmmode m\sbl H \else $m\sbl H$\fi}
\def\mch{\ifmmode m_{H^\pm} \else $m_{H^\pm}$\fi}
\def\mt{\ifmmode m_t\else $m_t$\fi}
\def\mc{\ifmmode m_c\else $m_c$\fi}
\def\mz{\ifmmode M_Z\else $M_Z$\fi}
\def\mw{\ifmmode M_W\else $M_W$\fi}
\def\mws{\ifmmode M_W^2 \else $M_W^2$\fi}
\def\mhs{\ifmmode m_H^2 \else $m_H^2$\fi}   
\def\mzs{\ifmmode M_Z^2 \else $M_Z^2$\fi}
\def\mts{\ifmmode m_t^2 \else $m_t^2$\fi}
\def\mcs{\ifmmode m_c^2 \else $m_c^2$\fi}
\def\mchs{\ifmmode m_{H^\pm}^2 \else $m_{H^\pm}^2$\fi}
\def\ztwo{\ifmmode Z_2\else $Z_2$\fi}
\def\zone{\ifmmode Z_1\else $Z_1$\fi}
\def\mtwo{\ifmmode M_2\else $M_2$\fi}
\def\mone{\ifmmode M_1\else $M_1$\fi}
\def\tb{\ifmmode \tan\beta \else $\tan\beta$\fi}
\def\xw{\ifmmode x\sub w\else $x\sub w$\fi}
\def\ch{\ifmmode H^\pm \else $H^\pm$\fi}
\def\lum{\ifmmode {\cal L}\else ${\cal L}$\fi}
\def\inpb{\ifmmode {\rm pb}^{-1}\else ${\rm pb}^{-1}$\fi}
\def\infb{\ifmmode {\rm fb}^{-1}\else ${\rm fb}^{-1}$\fi}
\def\epem{\ifmmode e^+e^-\else $e^+e^-$\fi}
\def\ppb{\ifmmode \bar pp\else $\bar pp$\fi}

\def\bsg{\ifmmode b\rightarrow s\gamma \else $b\rightarrow s\gamma$\fi}
 
\newskip\zatskip \zatskip=0pt plus0pt minus0pt
\def\matth{\mathsurround=0pt}

\def\atversim#1#2{\lower0.7ex\vbox{\baselineskip\zatskip\lineskip\zatskip
  \lineskiplimit 0pt\ialign{$\matth#1\hfil##\hfil$\crcr#2\crcr\sim\crcr}}}

\begin{document}
\rightline{\vbox{\halign{&#\hfil\cr
&SLAC-PUB-7279\cr
&September 1996\cr}}}
\vspace{0.8in} 
\Title{Searches for New Gauge Bosons at Future Colliders}
\bigskip
\Author{Thomas G. Rizzo\doeack}
\Address{\SLAC}
\bigskip
\begin{Abstract}
The search reaches for new gauge bosons at future hadron and lepton colliders 
are summarized for a variety of extended gauge models. Experiments at these 
energies will vastly improve over present limits and will easily discover a 
$Z'$ and/or $W'$ in the multi-TeV range. 
\end{Abstract}
\bigskip
%I added this vskip
\vskip1.0in
\begin{center}
To appear in the {\it Proceedings of the 1996 DPF/DPB Summer Study on New
 Directions for High Energy Physics-Snowmass96}, Snowmass, CO, 
25 June-12 July, 1996. 
\end{center}
%
%I added this bigskip...we may need vskips
\bigskip
\def\thefootnote{\fnsymbol{footnote}}
\setcounter{footnote}{0}
\newpage
\section{Introduction}
The discovery of new gauge bosons, $Z',W'$, would be the cleanest signature 
for new physics beyond the Standard Model(SM) and would 
signal an extension of the 
gauge group by an additional factor such as $U(1)$ or $SU(2)$. Present direct 
searches 
for such particles at the Tevatron{\cite {tev}} suggest that, if they exist, 
their masses are in excess of several hundreds of GeV. It is thus the role of 
future colliders to search for a new $Z'$ or $W'$ at or 
above the TeV scale. In this 
paper we provide an overview and comparison of the capability of future hadron 
and lepton machines to discover these particles. 

The search reach at a collider for new gauge bosons is somewhat model 
dependent due to the rather large variations in their 
couplings to the SM fermions which are 
present in extended gauge theories currently on the market in the literature. 
This implies that any overview of the subject is necessarily incomplete. 
Hence, we will be forced to limit ourselves to a few representative models. 
In what follows, we chose as examples the set of models recently discussed by 
Cvetic and Godfrey{\cite {rev}} so that we need to say very little here about 
the details of the coupling structure of each scenario. To be specific we 
consider ({\it i}) the $E_6$ effective rank-5 model(ER5M), which predicts a 
$Z'$ whose couplings depend on a single parameter 
$-\pi/2 \leq \theta \leq \pi/2$ (with models $\psi$, $\chi$, $I$, and $\eta$ 
denoting specific $\theta$ values); ({\it ii}) the Sequential Standard 
Model(SSM) 
wherein the new $W'$ and $Z'$ are just heavy versions of the SM particles (of 
course, this is not a true model in the strict sense but is commonly used as a 
guide by experimenters); ({\it iii}) the Un-unified Model(UUM), based on the 
group $SU(2)_\ell \times SU(2)_q \times U(1)_Y$, which has a 
single free parameter $0.24 \leq s_\phi \leq 1$; 
({\it iv}) the Left-Right Symmetric Model(LRM), based on the group 
$SU(2)_L \times SU(2)_R \times U(1)_{B-L}$, 
which also has a free parameter $\kappa=g_R/g_L$ of order unity which is just 
the ratio of the gauge couplings 
and, lastly, ({\it v}) the Alternative Left-Right Model(ALRM), based on the 
same extended group as the LRM but now arising from 
$E_6$, wherein the fermion assignments are modified in comparison to the LRM 
due to an ambiguity in how they are embedded in the {\bf 27} representation. 

In the case of a $W'$ we will restrict ourselves to the specific example of 
the LRM, \ie, $W_R$, although both the UUM and ALRM have interesting $W'$ 
bosons. 
The $W'$ in the UUM is quite similar to that of the SSM apart from its overall 
coupling strength and the size of its leptonic branching fraction. The $W'$ 
in the ALRM cannot be singly produced via the Drell-Yan mechanism since it 
carries non-zero lepton number and negative $R-$parity{\cite {physrep}}. 
In what follows $Z-Z'$ and $W-W'$ mixing effects will be ignored which 
is an excellent approximation for any new gauge bosons in the multi-TeV mass 
range.

\section{$Z'$ Searches at Hadron Colliders}

In what follows we will limit our discussion to the most conventional 
discovery channels involving $Z'$ and $W'$ decays to charged lepton pairs and 
charged leptons plus missing $E_t$, respectively. 
Regrettably, this leaves vast and fascinating territories untouched wherein, 
\eg, the new gauge boson decays to dijets, pairs of SM gauge bosons, 
or leptonic 
$W'$ decay modes not involving missing $E_t$. These possibilities require 
further study particularly at the LHC. 

\begin{figure}[htbp]
\centerline{
\psfig{figure=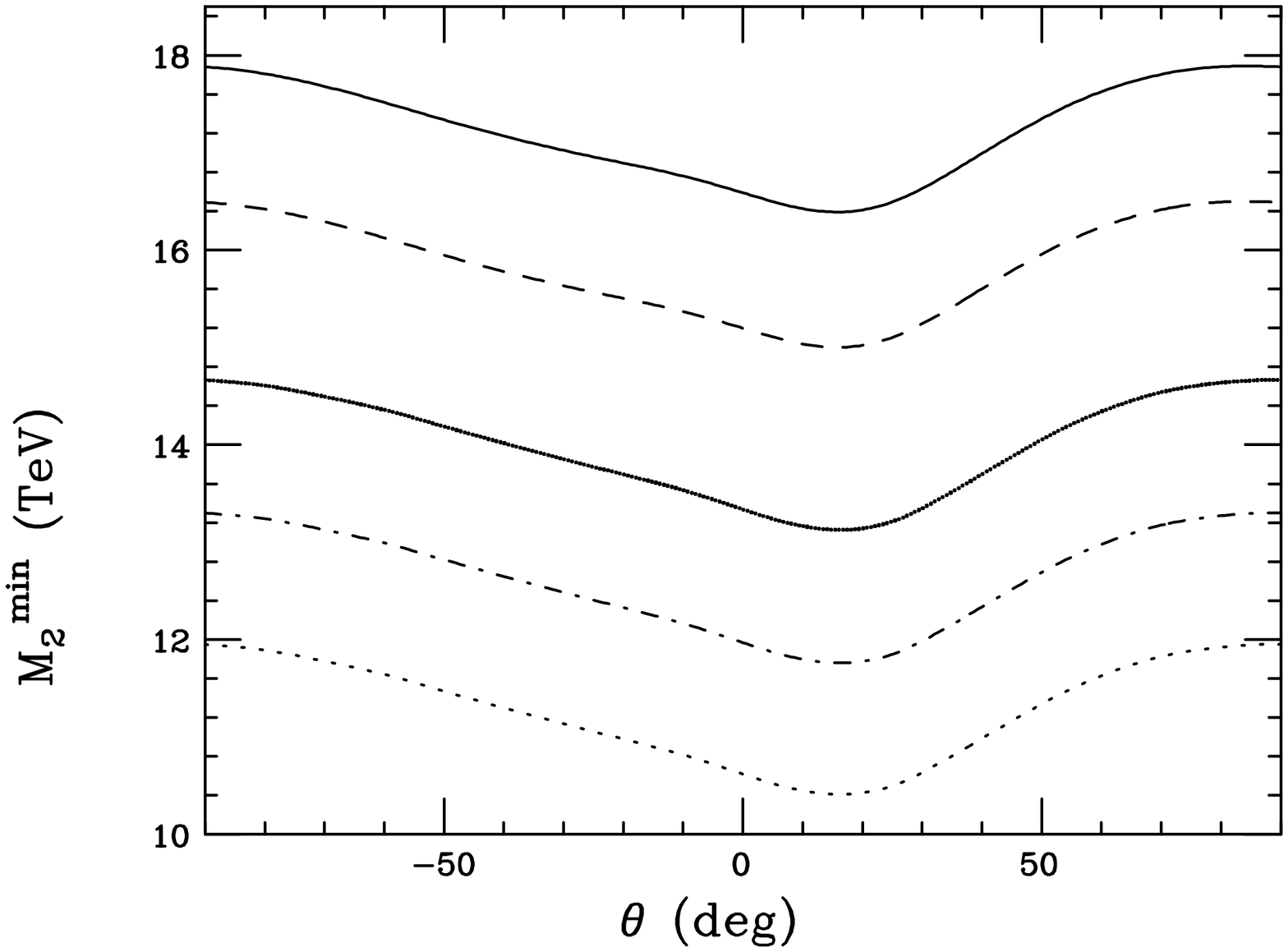,height=9.1cm,width=9.1cm,angle=-90}
\hspace*{-5mm}
\psfig{figure=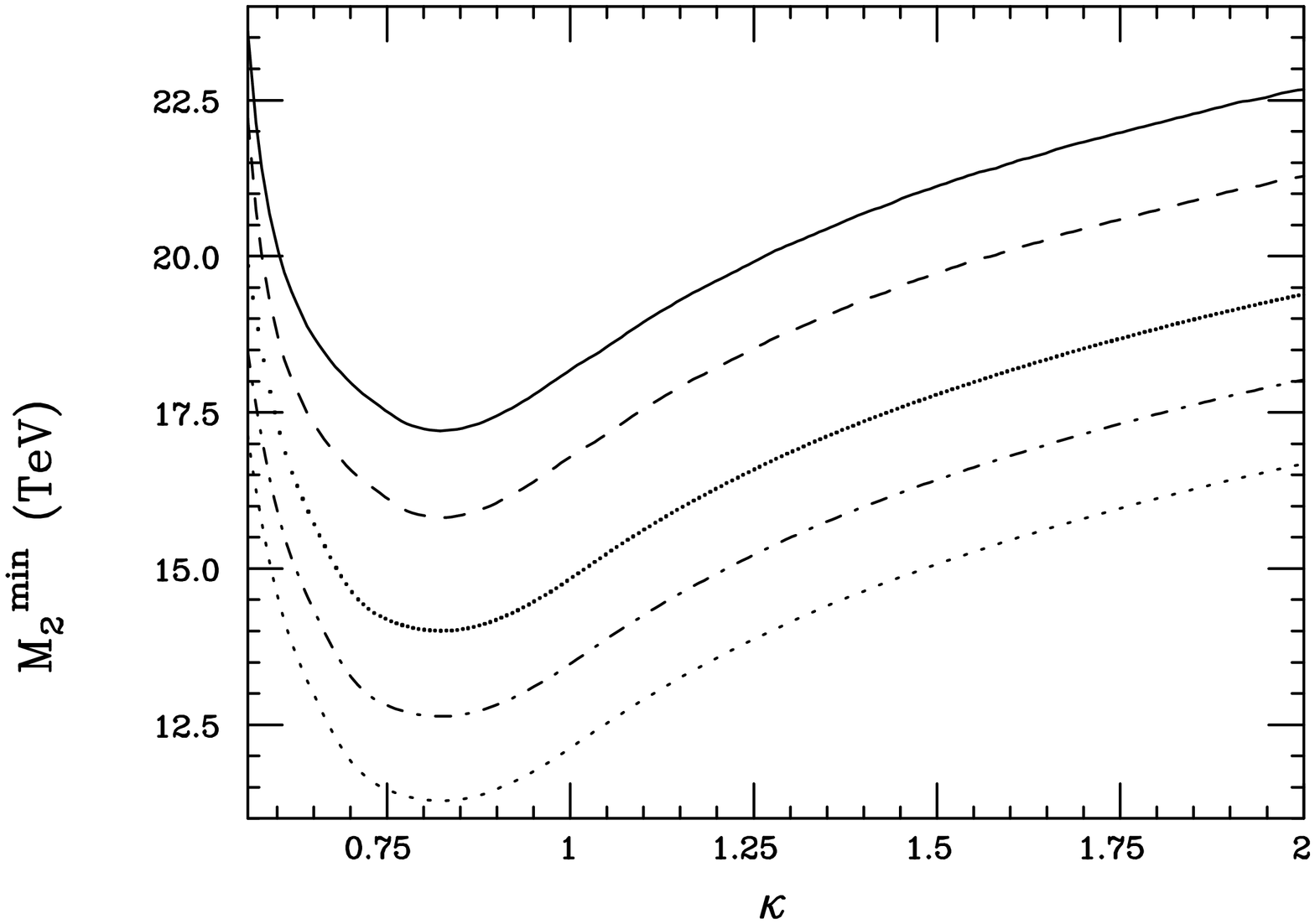,height=9.1cm,width=9.1cm,angle=-90}}
\vspace*{-0.6cm}
\caption{$Z'$ search reaches at the 60 TeV $pp$ collider(LSGNA) for $E_6$ 
models as a function of $\theta$ and the LRM as a function of $\kappa$. From 
bottom to top the curves correspond to integrated luminosities of 50, 100, 
200, 500 and 1000 $fb^{-1}$, respectively. MRSA$'$ parton densities 
are assumed.}
\label{figlhc}
\end{figure}

\begin{figure}[htbp]
\centerline{
\psfig{figure=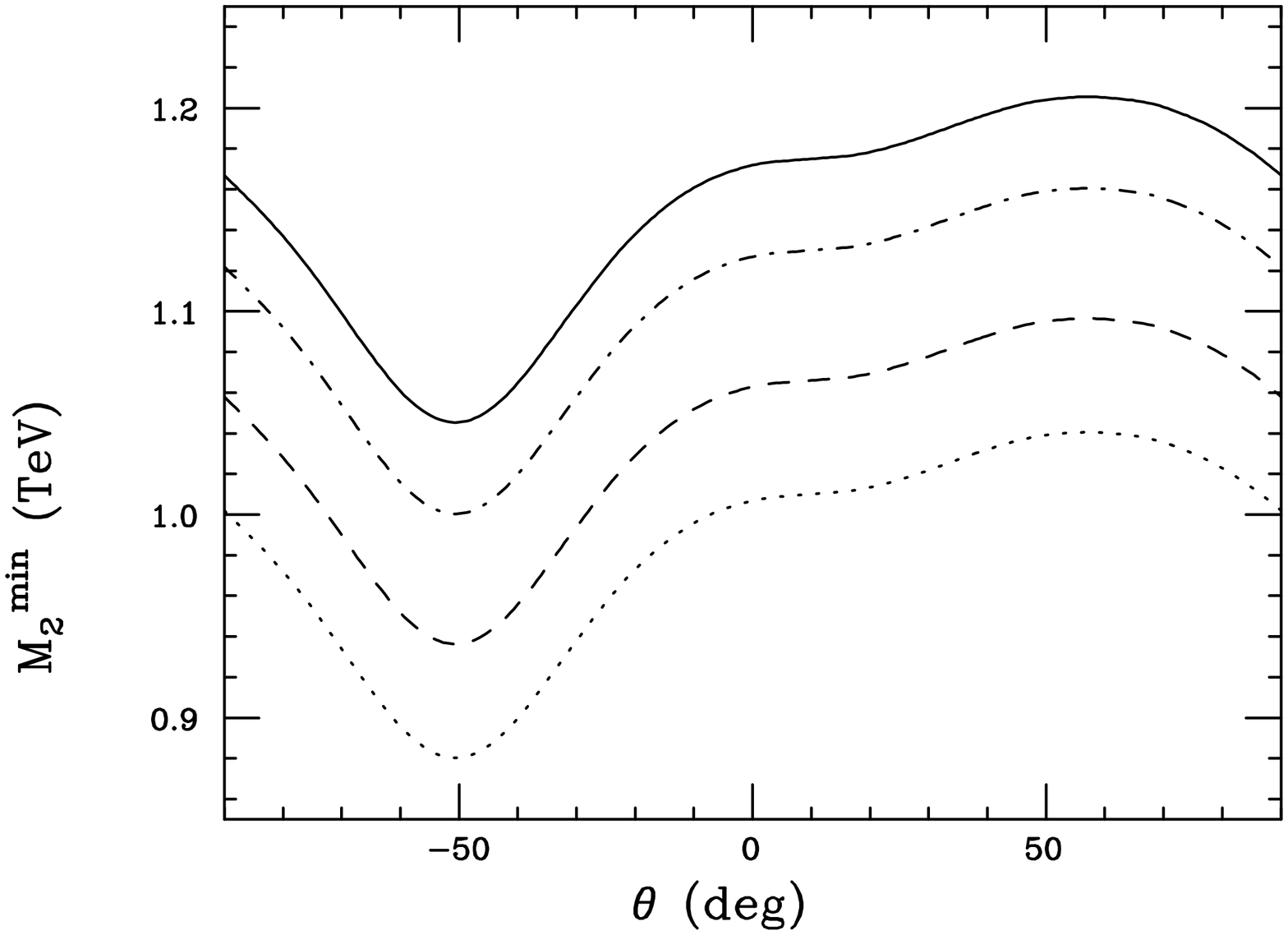,height=9.1cm,width=9.1cm,angle=-90}
\hspace*{-5mm}
\psfig{figure=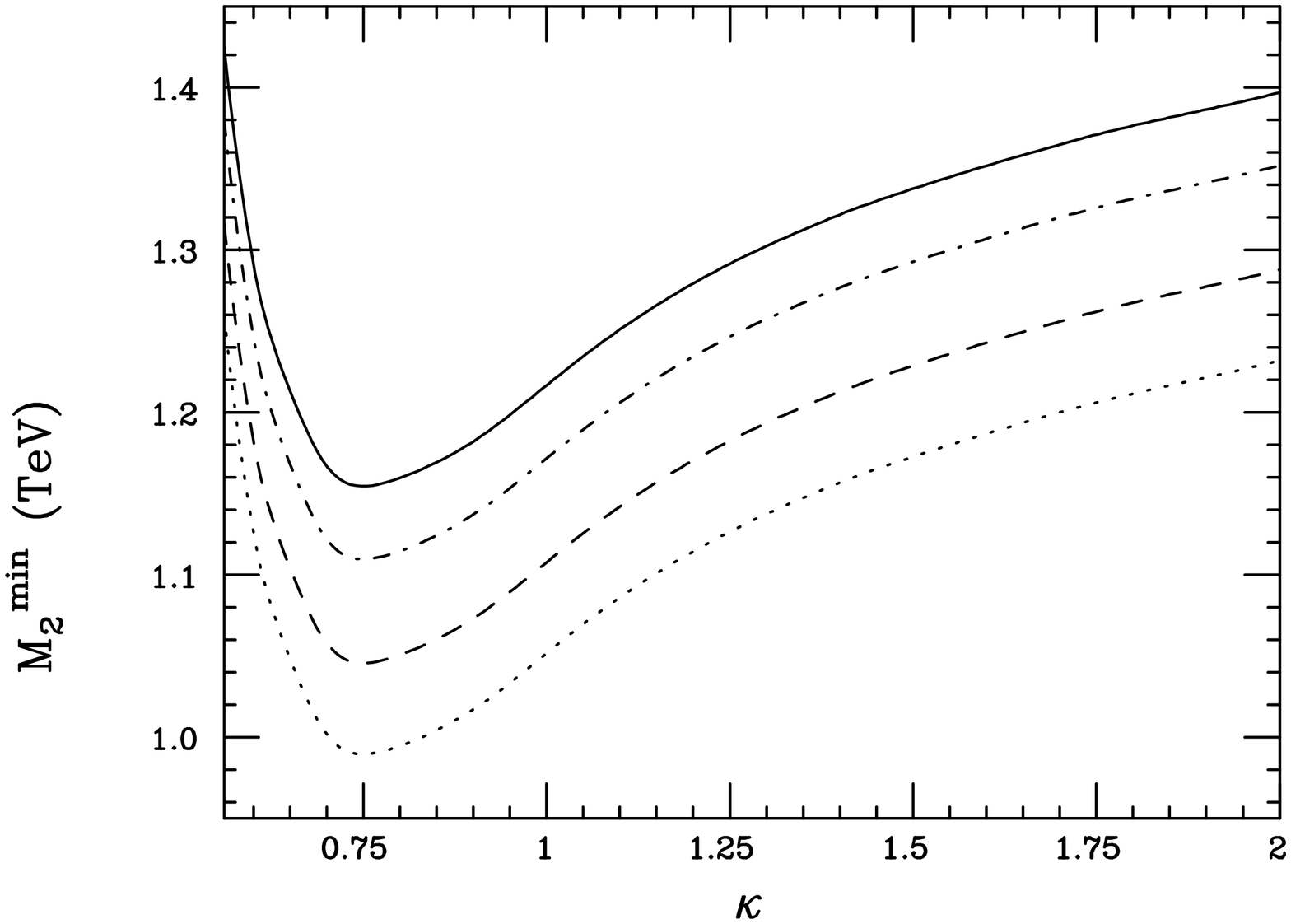,height=9.1cm,width=9.1cm,angle=-90}}
\vspace*{-0.6cm}
\caption{ Same as the previous figure, but now for the Tevatron running at 
2 TeV. From top to bottom the integrated luminosities are assumed to be 100, 
50, 20 and 10 $fb^{-1}$, respectively.}
\label{figtev}
\end{figure}

Both $Z'$ and $W'$ search reaches are conventionally obtained using the 
narrow width 
approximation with some additional corrections to account for detector 
acceptance's($A$) and efficiencies($\epsilon$). In this case the number of 
expected events($N$) is simply the product 
$N=\sigma B_l A \epsilon {\cal L}$, where $\sigma$ is the 
production cross section, $B_l$ is the leptonic branching fraction and 
${\cal L}$ is the machine's integrated luminosity. 
A $5\sigma$ signal is assumed to be given by 10 signal events with no 
background; this is logically consistent since an extremely narrow peak in the 
dilepton mass can have only an infinitesimal background underneath it. 
Detailed detector simulations for both the Tevatron and 
LHC{\cite {wulzlowe}} validate this 
approximation as a good estimator of the true search reach at least for the 
more `traditional' models. (The reader should be reminded to be careful when 
employing this approximation in all models since the $Z'$ may not always be 
sufficiently narrow and Drell-Yan continuum backgrounds may become relevant.) 
In the $Z'$ case, 
we need only know the various fermionic couplings for a fixed value of the 
$Z'$ mass to obtain $\sigma$. Traditionally, one also assumes that the 
$Z'$ can {\it only} decay to pairs of SM fermions in order to obtain 
$B_l$. It is important to note that in many models, 
where the $Z'$ can also decay to exotic fermions and/or 
SUSY particles this {\it overestimates} $B_l$ and, thus, the search reach. In 
obtaining our results for 10 signal events we combine both the electron and 
muon decay channels. 
With these assumptions, Figures~\ref{figlhc} and ~\ref{figtev} show the 
discovery reaches of the 60 TeV $pp$ (LSGNA) collider and TeV33 for the $Z'$ 
bosons of both the ER5M and the LRM, 
while Table~\ref{$Z'$ lhc} shows the summary of results for the other models as 
well as for the LHC and the higher energy 200 TeV (PIPETRON) colliders. The 
corresponding figures for the LHC can be found in Ref.{\cite {bsm}}. Here we 
see that TeV33 will allow us to approach the 1 TeV mass scale for $Z'$ bosons 
for the first time. Note that 
in the case of the 60 and 200 TeV machines the higher $q\bar q$ 
luminosities in the $p\bar p$ mode leads to a significantly 
greater ($\simeq 30-50\%$) search reach.

%%%%% A Table
%%
\begin{table}[htpb]
\centering
\begin{tabular}{lcccccc}
\hline
\hline
Model & LHC&60 TeV ($pp$)&60 TeV ($p\bar p$)&200 TeV ($pp$)&
200 TeV ($p\bar p$)&TeV33 \\
\hline
$\chi$  & 4.49 & 13.3 & 17.5 & 43.6 & 63.7 & 1.00    \\
$\psi$  & 4.14 & 12.0 & 17.1 & 39.2 & 62.3 & 1.01    \\
$\eta$  & 4.20 & 12.3 & 17.9 & 40.1 & 64.8 & 1.03    \\
I       & 4.41 & 12.9 & 15.2 & 42.1 & 56.0 & 0.88    \\
SSM     & 4.88 & 14.4 & 20.6 & 45.9 & 68.7 & 1.10     \\
ALRM    & 5.21 & 15.0 & 22.5 & 49.9 & 74.7 & 1.15     \\
LRM     & 4.52 & 13.5 & 18.9 & 43.2 & 64.6 & 1.05     \\
UUM     & 4.55 & 13.7 & 19.7 & 43.5 & 65.1 & 1.08     \\
        & & & & & & \\
Hit     & 0.33 & 1.5 & 1.8 & 4.9 & 6.3 & 0.05 \\
\hline
\hline
\end{tabular}
\caption{$Z'$ search reaches at hadron colliders in TeV. For the LRM, 
$\kappa=1$ is assumed while for the UUM, we take $s_\phi=0.5$. Decays to only 
SM fermions is assumed. The luminosities of the Tevatron, LHC, 60 TeV and 
200 TeV colliders are assumed to be 10, 100, 100 and 1000 $fb^{-1}$, 
respectively. The last line in the Table is the approximate reduction in 
reach in TeV due to a decrease in $B_l$ by a factor of 2.}
\label{$Z'$ lhc}
\end{table}

If the above estimate of the leptonic branching fraction is wrong, how badly 
are the reaches affected? To get a feeling for this, consider reducing the 
value of $B_l$ by a factor of two from the naive estimate given by decays to 
only SM fermion pairs. (In the $E_6$ case, this roughly corresponds to 
allowing the $Z'$ to decay into SUSY partners as well as the exotic fermions 
with some phase space suppression{\cite {physrep}}.) 
Semi-quantitatively, the reduction in reach for each collider is found to be 
roughly model independent and approximate results are given in the last line of 
Table~\ref{$Z'$ lhc}. As can be seen from these values the `hit' taken can be 
significant in some cases. However, unless $B_l$ is very much smaller than the 
naive estimate it is clear that the multi-TeV mass range will remain 
easily accessible to future hadron colliders. 

\vspace*{-0.5cm}
\nn
\vspace*{0.1mm}
\vspace*{-0.5cm}
\nn
\vspace*{0.1mm}

\section{$W'$ Searches at Hadron Colliders}

Unlike the $Z'$ case, the corresponding $W_R$ searches 
via the Drell-Yan process have many subtleties even when we assume that the 
missing $E_t$ mode is accessible and dominant. The canonical search assumes 
that the $q'\bar qW_R$ 
production vertex has SM strength, implying ({\it i}) $\kappa=1$ and 
({\it ii}) $|V_{L_{ij}}|=|V_{R_{ij}}|$, \ie, the elements of the RH CKM mixing 
matrix, $V_R$, are the same as $V_L$, and, as in the $Z'$ case, ({\it iii})  
that the $W_R$ leptonic branching fraction is given by its decay to SM 
fermions only. Of course violations of 
assumptions ({\it i}) and ({\it iii}) are easily accounted for in a manner 
similar to the $Z'$ case discussed above. 
If assumption ({\it ii}) is invalid, a significant search reach degradation 
can easily occur as a result 
of modifying the weight of the various parton luminosities which enter into 
the calculation of the production cross section. At the $pp$ colliders such 
as the LHC, we do not expect that surrendering ({\it ii}) 
will cost us such a very large penalty since the $W_R$ production process 
already occurs 
through the annihilation of sea$\times$valence quarks. On the otherhand, $W_R$  
production is a valence$\times$valence process at the $p\bar p$ colliders 
such as the Tevatron so we might anticipate a more significant reach 
reduction in this case.

\begin{figure}[htbp]
\centerline{
\psfig{figure=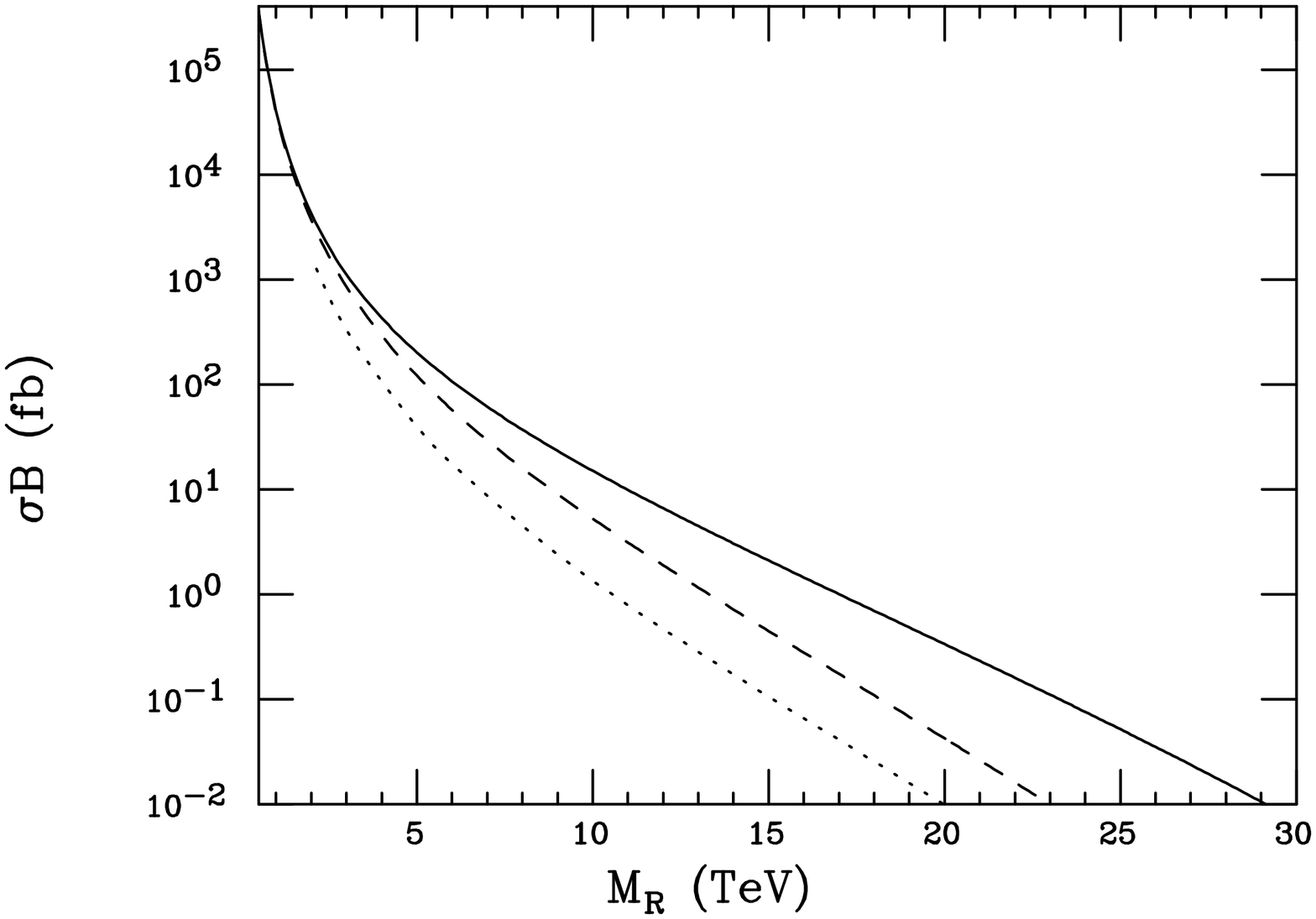,height=9.1cm,width=9.1cm,angle=-90}
\hspace*{-5mm}
\psfig{figure=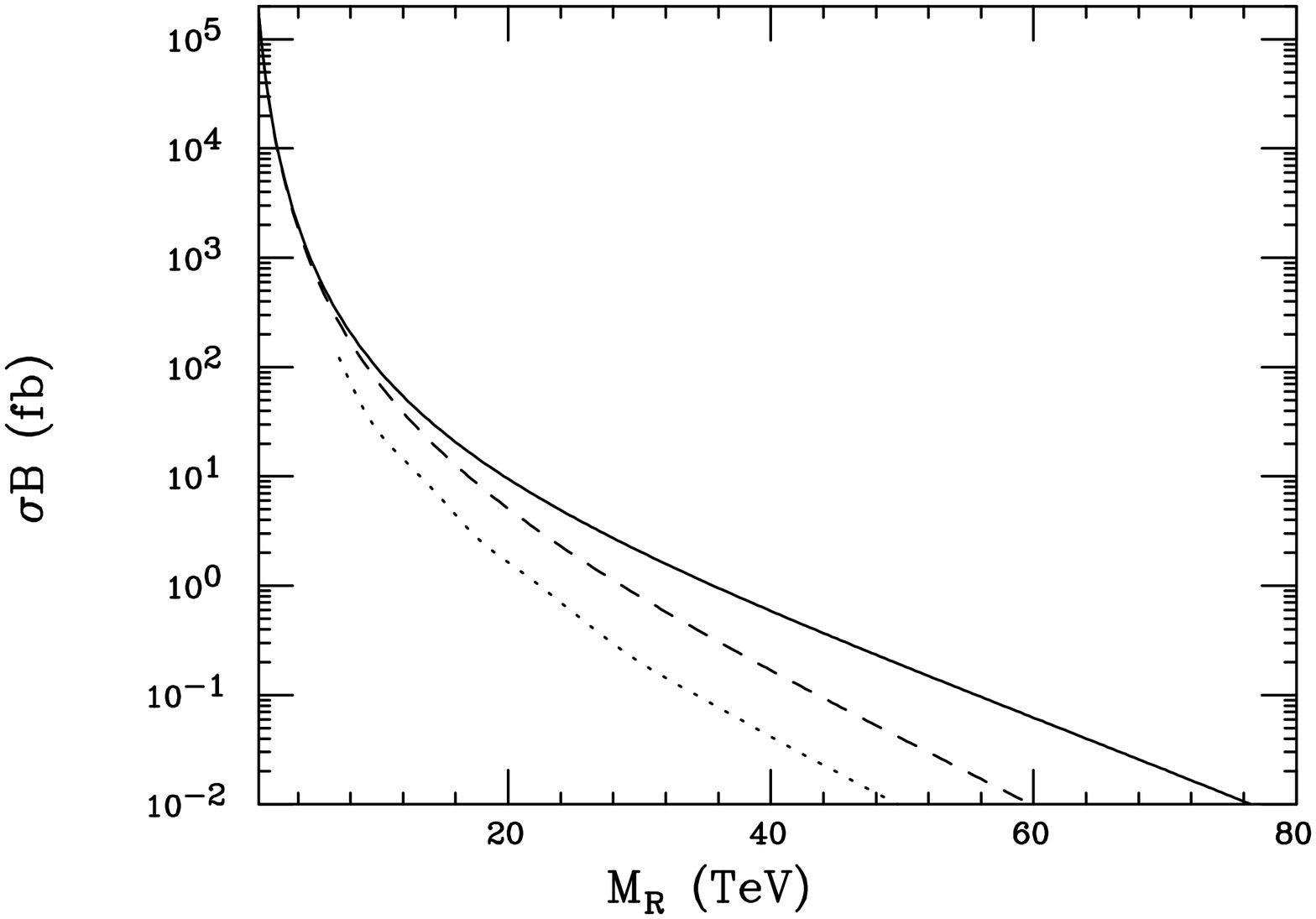,height=9.1cm,width=9.1cm,angle=-90}}
\vspace*{-0.6cm}
\caption{$W_R$ production cross sections for $\kappa=1$ at the 60 and 200 
TeV colliders. $B_l$ is assumed to be given by decays to the SM fermions only. 
The solid(dashed) curve corresponds to $p\bar p(pp)$ collisions 
with $V_L=V_R$ while the dotted curve corresponds to the lowest cross section 
in either case due to the most pessimistic choice of the $V_R$ mixing matrix 
elements.}
\label{figwlhc}
\end{figure}

Fig.2 of Ref.{\cite {bsm}} summarizes the $W_R$ search 
reach situation at both the Tevatron and the LHC where the narrow width 
approximation has been employed. In particular this figure 
shows that the reduction of reach at the LHC due to variations in $V_R$ is 
rather modest whereas it is far more significant at the Tevatron. 
Figure~\ref{figwlhc} compares the $W_R$ production rates at the 60 and 200 TeV 
colliders for both $pp$ and $p\bar p$ modes assuming $\kappa=1$. In both cases 
we see that the maximum reach degradation resulting from variations in $V_R$ 
is far more severe in the $p\bar p$ than $pp$ mode. For both the 60 and 200 TeV 
colliders the search reach is $\simeq 25\%$ higher in the case of $p\bar p$. 
It is also interesting to 
compare the rates expected for the $p\bar p$ and $pp$ modes for a fixed value 
of $\sqrt s$ and $W_R$ mass($M_R$). For example, at the 60(200) TeV machine 
and $M_R=12(60)$ TeV, the production rates are found to be 6.62, 1.90, and 
0.4(0.588, 0.168 and 0.04) $fb$ in the $p\bar p$, $pp$, and $V_R$ `worst case' 
modes. 

\begin{figure}[htbp]
\centerline{
\psfig{figure=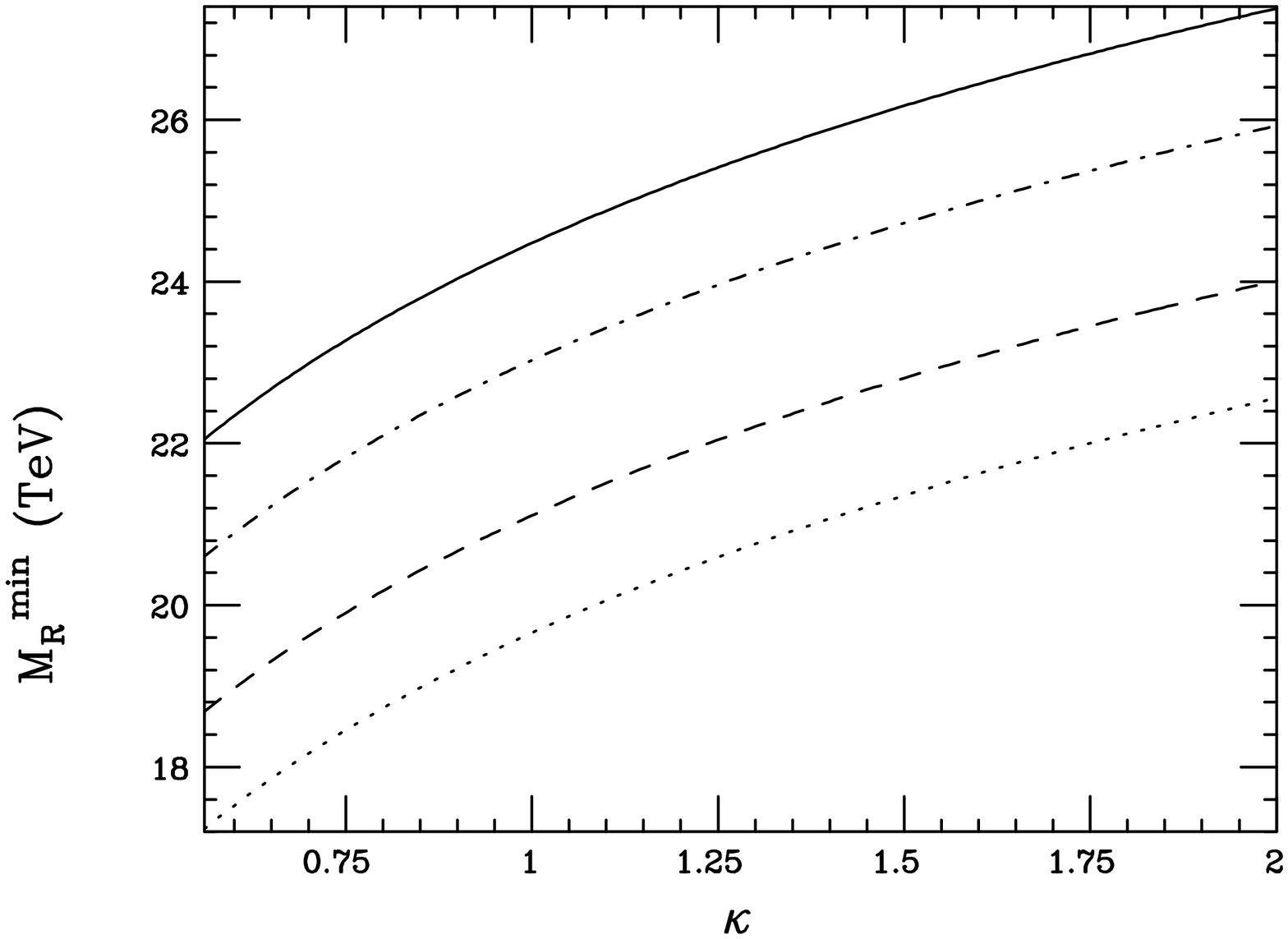,height=9.1cm,width=9.1cm,angle=-90}
\hspace*{-5mm}
\psfig{figure=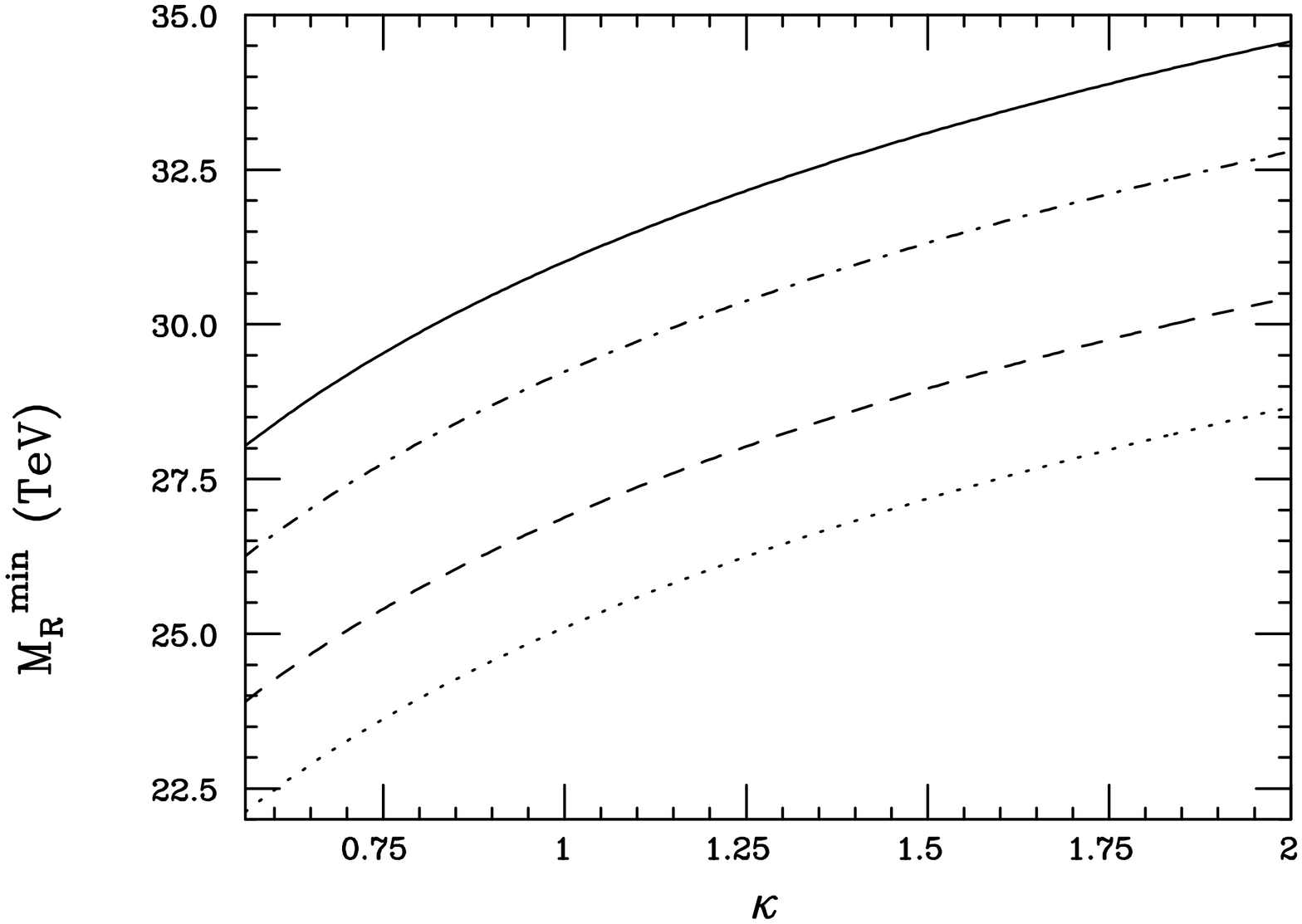,height=9.1cm,width=9.1cm,angle=-90}}
\vspace*{-0.6cm}
\caption{$W_R$ search reaches at the 60 TeV LSGNA collider in the $pp$(left) 
and $p\bar p$(right) modes as functions of $\kappa$ assuming $V_R=V_L$. 
From top to bottom the curves correspond to integrated luminosities of 1000, 
500, 200 and 100 $fb^{-1}$, respectively.}
\label{figw2lhc}
\end{figure}

Assuming $V_R=V_L$ for the 
60 TeV collider, Figure~\ref{figw2lhc} compares the $\kappa$ dependence of 
the reach for both the $pp$ and $p\bar p$ modes for different integrated 
luminosities. Table~\ref{$W_R$ lhc} summarizes all of our results for $W_R$ 
search reaches at various colliders.

\vspace*{-0.5cm}
\nn
\vspace*{0.1mm}
\vspace*{-0.5cm}
\nn
\vspace*{0.1mm}
%

%%%%% A SECOND Table
%%
\begin{table}[htpb]
\centering
\begin{tabular}{lcc}
\hline
\hline
Machine  & $V_L=V_R$  & $V_R$ (WC) \\
\hline
TeV33               &  1.2 & $\simeq 0.5$\\
LHC                 &  5.9 &  5.1 \\
60 TeV ($pp$)       & 19.7 & $\simeq 16$ \\
60 TeV ($p\bar p$)  & 25.1 & $\simeq 16$ \\
200 TeV ($pp$)      & 64.7 & $\simeq 52$ \\
200 TeV ($p\bar p$) & 82.9 & $\simeq 52$ \\
\hline
\hline
\end{tabular}
\caption{$W_R$ search reaches of hadron colliders in the missing energy mode 
in TeV. 
$\kappa=1$ and decays to only SM fermions is assumed. WC(worst case) 
refers to the set of 
$V_R$ elements that yield the lowest production cross section. The 
luminosities are as in the previous Table.}
\label{$W_R$ lhc}
\end{table}
%%
%%%%%

\section{$Z'$ Searches at Lepton Colliders}

It is more than likely that a $Z'$ will be too massive to be produced directly 
at the first generation of new lepton colliders. Thus searches at such 
machines will be indirect and will consist of 
looking for deviations in the predictions of the SM in as many observables as 
possible. Layssac \etal {\cite {rev}} have shown that the deviations in the 
leptonic observables due to the existence of a $Z'$ are rather unique. Since 
the $Z'$ is not directly produced, lepton collider searches are insensitive to 
the decay mode assumptions that we had to make in the case of hadron colliders. 
In the analysis presented here we consider the following standard set of 
observables: $\sigma_{f}$, $A_{FB}^{f}$, $A_{LR}^{f}$, $A_{pol}^{FB}(f)$ 
where $f$ labels the fermion in the final state and, special to the case of 
the tau, $<P_\tau>$ and $P_\tau^{FB}$. Note 
that beam polarization plays an important role in this list of observables, 
essentially doubling its length. 

In 
this paper we present a preliminary analysis wherein charged leptons as well as 
$b-$, $c-$, and $t-$quarks are considered simultaneously in obtaining the 
discovery reach. The basic approach follows that of Hewett and 
Rizzo{\cite {hr}} and is outlined in the review of Cvetic and 
Godfrey{\cite {rev}}, but now includes angular cuts, initial state 
radiation(ISR) in the $e^+e^-$ case but ignored for $\mu^+\mu^-$ collisions at 
the Large Muon Collider(LMC), finite identification efficiencies, systematics 
associated with luminosity and beam polarization($P$) uncertainties.  
For $e^+e^-$ colliders we take $P=90\%$ while for the LMC we can trade off a 
smaller effective $P$ through modifications{\cite {lmc}} in the 
integrated luminosity. The angular cuts applied in all cases were assumed to 
be the same. 
Generically we find that ISR lowers the search reach by $15-20\%$ while finite 
beam polarization increases the reach by $15-80\%$ depending on the specific 
model and the machine energy, \ie, the increase is smaller at larger values of 
$\sqrt s$.

\vspace*{-0.5cm}
\nn
\begin{figure}[htbp]
\centerline{
\psfig{figure=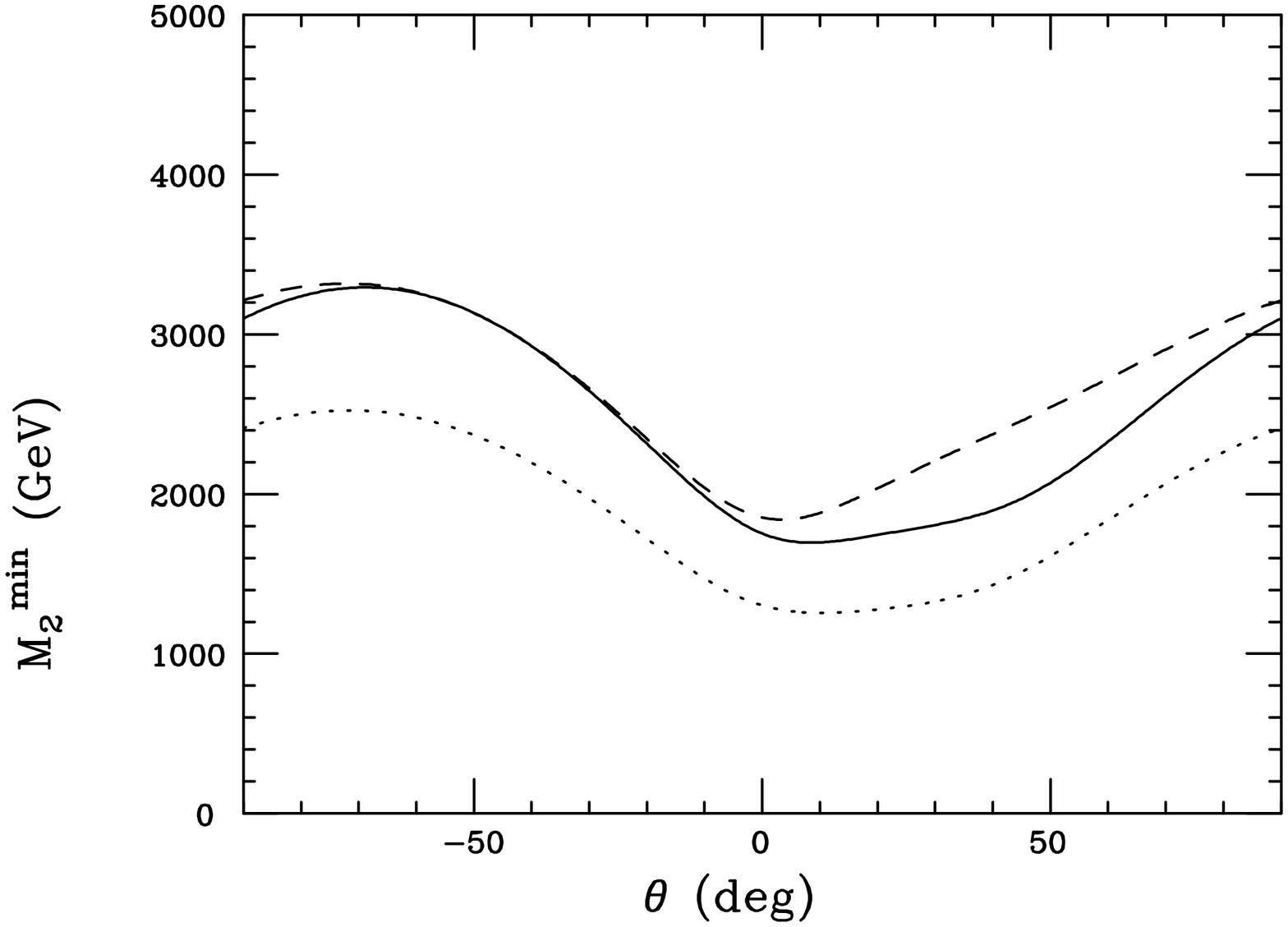,height=9.1cm,width=9.1cm,angle=-90}
\hspace*{-5mm}
\psfig{figure=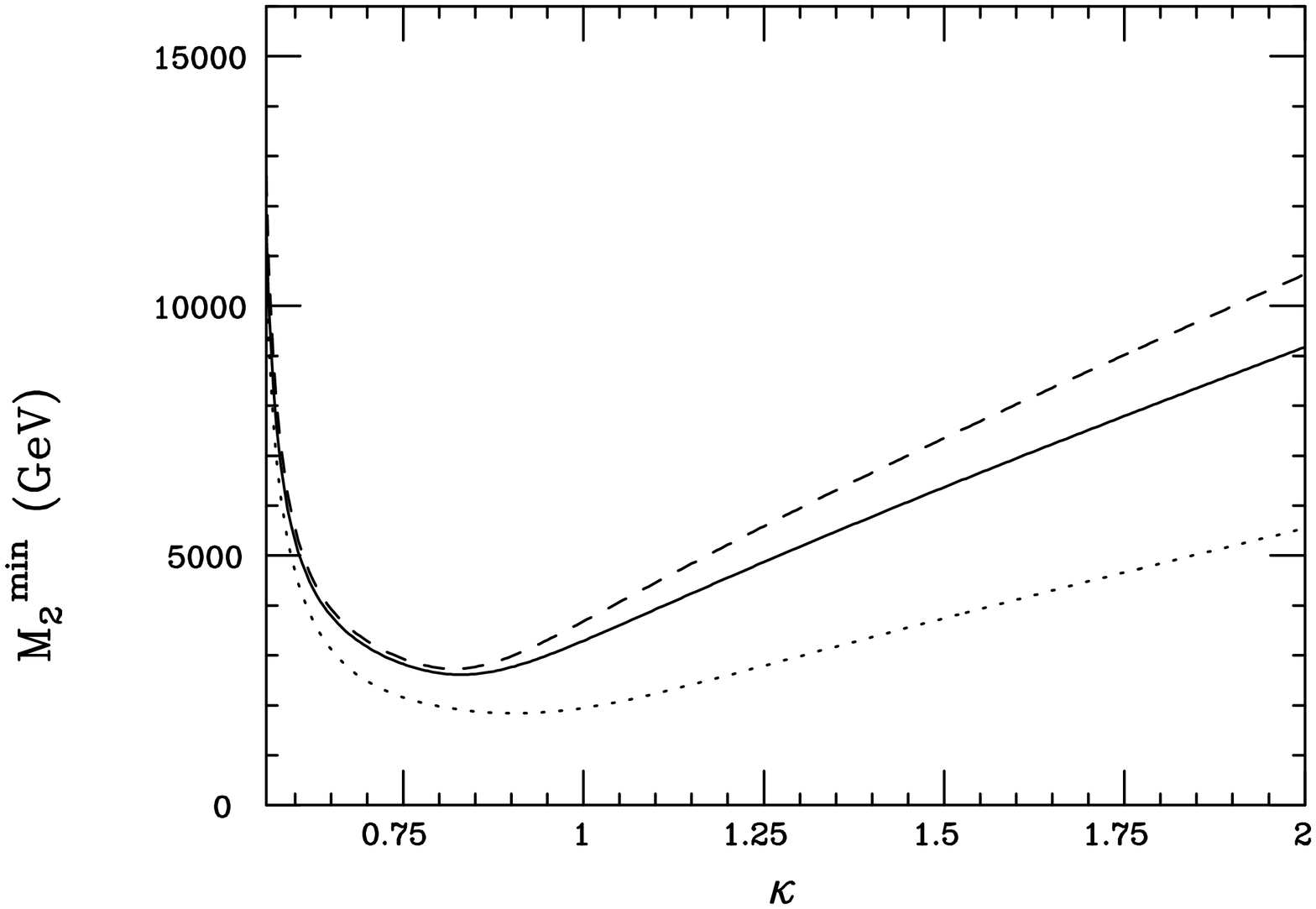,height=9.1cm,width=9.1cm,angle=-90}}
\vspace*{-0.6cm}
\caption{Indirect $Z'$ search reaches at the 500 GeV NLC for $E_6$ models as 
a function of $\theta$ and the LRM as a function of $\kappa$ including initial 
state radiation. The dotted(solid, dashed) curve corresponds to the values 
obtained using leptonic(leptonic plus $b-$quark, all) observables. A luminosity 
of 50 $fb^{-1}$ has been assumed.}
\label{fignlc}
\end{figure}
\vspace*{0.1mm}
\vspace*{-0.5cm}
\nn
\begin{figure}[htbp]
\centerline{
\psfig{figure=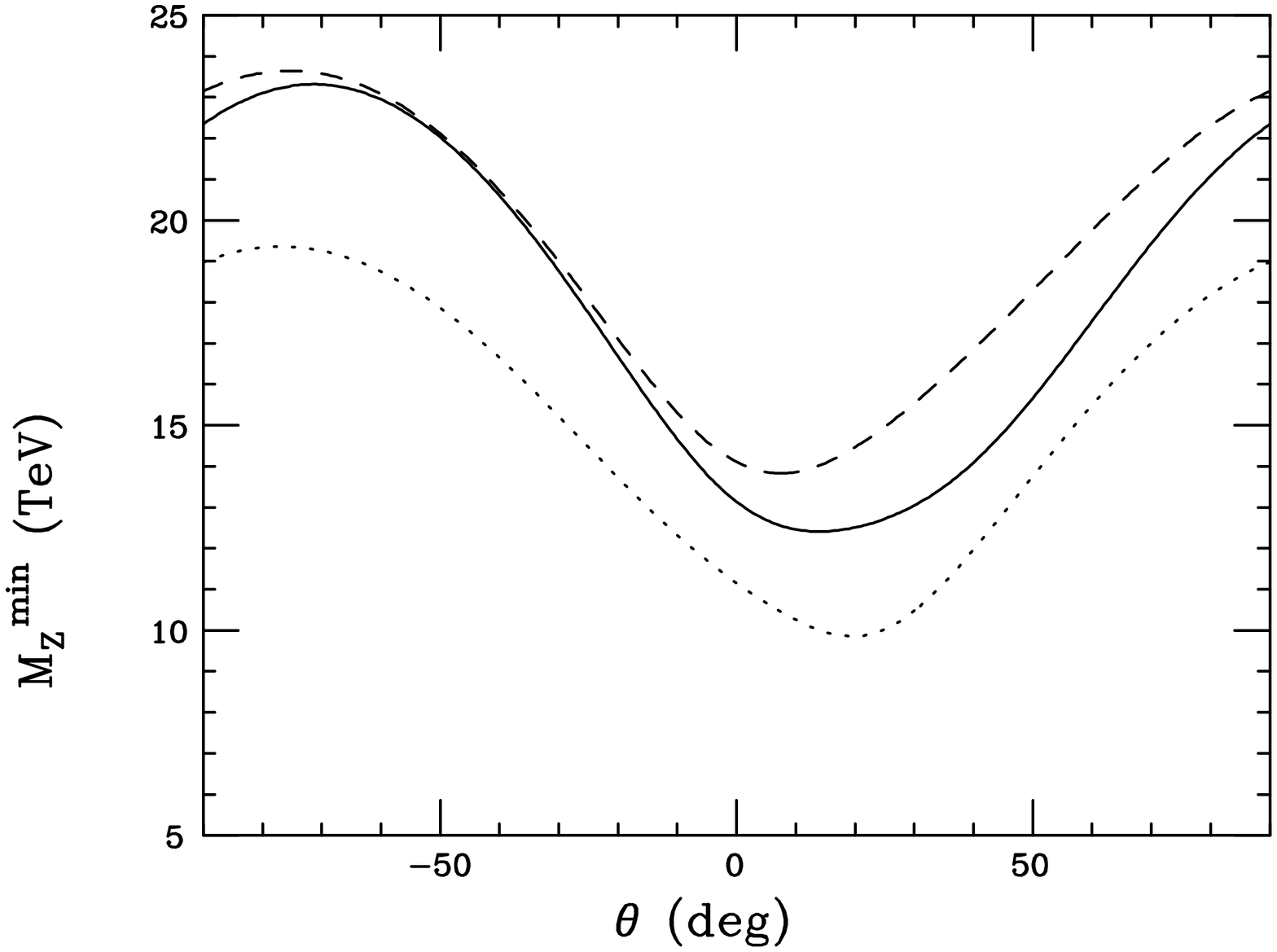,height=9.1cm,width=9.1cm,angle=-90}
\hspace*{-5mm}
\psfig{figure=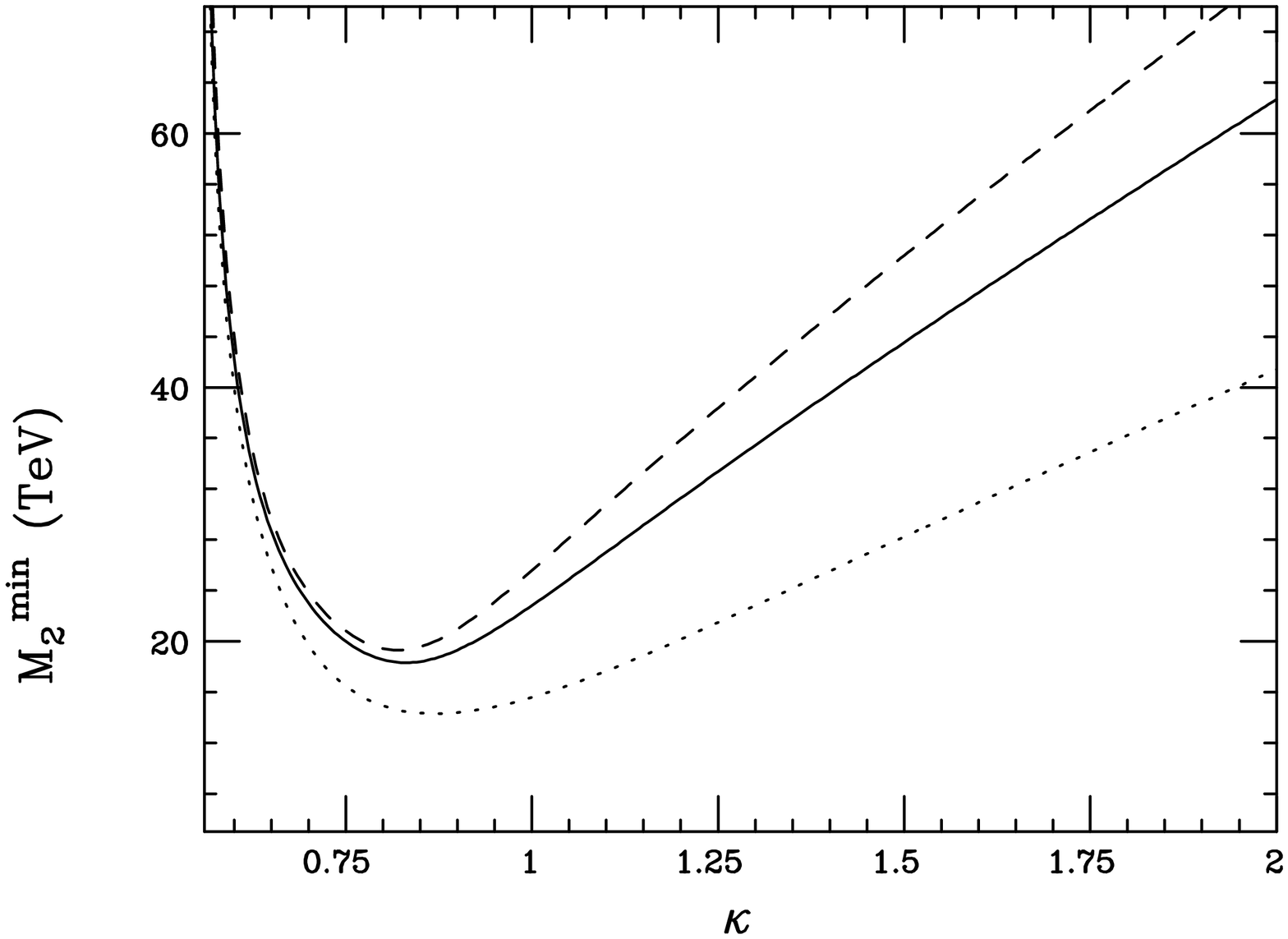,height=9.1cm,width=9.1cm,angle=-90}}
\vspace*{-0.6cm}
\caption{Same as the previous figure but now for the 5 TeV NNLC assuming an 
integrated luminosity of 1000 $fb^{-1}$.}
\label{fignnlc}
\end{figure}
\vspace*{0.1mm}
%

%%%%% Another Table
%%
\begin{table}[htpb]
\centering
\begin{tabular}{lccccc}
\hline
\hline
Model  & NLC500  &NLC1000 &NLC1500 &NNLC 5 TeV& LMC 4 TeV \\
\hline
$\chi$  & 3.21 & 5.46 & 8.03 & 23.2 & 18.2 \\
$\psi$  & 1.85 & 3.24 & 4.78 & 14.1 & 11.1 \\
$\eta$  & 2.34 & 3.95 & 5.79 & 16.6 & 13.0 \\
I       & 3.17 & 5.45 & 8.01 & 22.3 & 17.5 \\
SSM     & 3.96 & 6.84 & 10.1 & 29.5 & 23.2 \\
ALRM    & 3.83 & 6.63 & 9.75 & 28.4 & 22.3\\
LRM     & 3.68 & 6.28 & 9.23 & 25.6 & 20.1 \\
UUM     & 4.79 & 8.21 & 12.1 & 34.7 & 27.3 \\
\hline
\hline
\end{tabular}
\caption{Indirect $Z'$ search reaches of lepton colliders in TeV employing 
all observables. The integrated luminosities of the NLC500, NLC1000, NLC1500, 
NNLC and LMC are assumed to be 50, 100, 100, 1000 and 1000 $fb^{-1}$, 
respectively.}
\label{$Z'$ nlc}
\end{table}

Figures~\ref{fignlc} and ~\ref{fignnlc} display  sample results of this 
analysis at the 500 GeV NLC  and 5 TeV Next-to-Next Linear Collider(NNLC) for 
a $Z'$ of either the ER5M or LRM type. In particular, these plots show 
how the introduction of additional observables associated first with $b$ and 
then with $c$ and $t$ lead to an increased reach. Note that the inclusion of 
$c$ and $t$ in comparison to the leptons plus $b$ case leads to only a rather 
mild increase in the reach for the $E_6$ case with a somewhat larger result on 
the average for the LRM. One reason for this is that the $Q=2/3$ quarks have 
vanishing vectorial couplings for all values of the parameter $\theta$ and 
completely decouple from the $Z'$ in the case of model $I$ (which corresponds 
to $\theta \simeq -52.24^{\circ}$) so that there is 
no additional sensitivity obtained in this case when the $c$ and $t$ are 
included. 
Table~\ref{$Z'$ nlc} summarizes our results for the search reaches of the 
various colliders for all of the above models. It is interesting to note that 
for the LMC the lack of significant ISR and the smaller 
polarization/luminosity are found 
to essentially cancel numerically in their affect on the $Z'$ search reach.

In principle the NLC can be run in the polarized $e^-e^-$ collision mode with a 
luminosity comparable to that for $e^+e^-$. Since both $e^-$ beams are 
polarized, the {\it effective} polarization is larger and, due to the large 
Moller cross section, there is significant sensitivity to the existence of a 
$Z'${\cite {fc}}. Unfortunately, an analysis of this situation including the 
effects of ISR is not yet available but a preliminary study by 
Cuypers{\cite {fc}} indicates that the {\it ratio} of search reaches in the 
$e^+e^-$ and $e^-e^-$ modes is stable under the modifications induced by ISR. 
We thus repeat the previous $e^+e^-$ analysis neglecting ISR and also 
perform the 
complementary $e^-e^-$ analysis with the same cuts, efficiencies \etc ~and then 
take the ratio of the resulting reaches for a given extended gauge model. 
The results of this analysis for NLC500 are shown in 
Table~\ref{$e^-e^-$}. Here we see that in general the $e^-e^-$ reach is 
superior to that obtained in the $e^+e^-$ mode when only the leptonic 
final states are used, 
consistent with the results obtained in Ref.{\cite {fc}}. However, as soon as 
one adds the additional information from the quark sector, $e^+e^-$ regains 
the lead in terms of $Z'$ mass reach. Combining the leptonic and quark data 
together in the $e^+e^-$ case always results in a small value for the ratio. 
Of course, once the anxiously awaited $e^-e^-$ analysis including ISR becomes 
available we need to verify these results directly.

%%%%% Another Table
%%
\begin{table}[htbp]
\centering
\begin{tabular}{lccc}
\hline
\hline
Model  & $\ell$  & $\ell+b$ & $\ell+b,c,t$ \\
\hline
$\chi$  &  1.10 & 0.900 & 0.896 \\
$\psi$  &  1.20 & 0.711 & 0.673 \\
$\eta$  &  1.07 & 0.813 & 0.650 \\
I       &  1.06 & 0.813 & 0.813 \\
SSM     &  1.30 & 0.752 & 0.667 \\
ALRM    &  1.20 & 1.12  & 0.909 \\
LRM     &  1.02 & 0.483 & 0.432 \\
UUM     & 0.891 & 0.645 & 0.496 \\
\hline
\hline
\end{tabular}
\caption{Ratio of $e^-e^-$ to $e^+e^-$ indirect $Z'$ search reaches at a 500 
GeV NLC with an integrated luminosity of 50 $fb^{-1}$ in either collision 
mode. ISR 
has been ignored. The columns label the set of the final state fermions used in 
the $e^+e^-$ analysis.}
\label{$e^-e^-$}
\end{table}

\section{Acknowledgements}

The author would like to thanks S. Godfrey, J. Hewett, K. Maeshima and 
S. Riemann for discussions related to this work.

%
%%%%%%%%%%%%%%%%%%--- References
%%%%%%%%%%%%%%%%%%%%%%%%%%%%%%%%%%%%%%%%%%%%%%%%%%%%%%%
\def\MPL #1 #2 #3 {Mod.~Phys.~Lett.~{\bf#1},\ #2 (#3)}
\def\NPB #1 #2 #3 {Nucl.~Phys.~{\bf#1},\ #2 (#3)}
\def\PLB #1 #2 #3 {Phys.~Lett.~{\bf#1},\ #2 (#3)}
\def\PR #1 #2 #3 {Phys.~Rep.~{\bf#1},\ #2 (#3)}
\def\PRD #1 #2 #3 {Phys.~Rev.~{\bf#1},\ #2 (#3)}
\def\PRL #1 #2 #3 {Phys.~Rev.~Lett.~{\bf#1},\ #2 (#3)}
\def\RMP #1 #2 #3 {Rev.~Mod.~Phys.~{\bf#1},\ #2 (#3)}
\def\ZP #1 #2 #3 {Z.~Phys.~{\bf#1},\ #2 (#3)}
\def\IJMP #1 #2 #3 {Int.~J.~Mod.~Phys.~{\bf#1},\ #2 (#3)}

\end{document}